\documentclass[useAMS,usenatbib]{mn2e}
\usepackage{epsfig}
\usepackage{amsmath}
\usepackage{rotating}
\usepackage{color}

\newcommand{\be}{\begin{equation}}
\newcommand{\beq}{\begin{equation}}
\newcommand{\ba}{\begin{eqnarray}}
\newcommand{\ee}{\end{equation}}
\newcommand{\eeq}{\end{equation}}
\newcommand{\ea}{\end{eqnarray}}

\newcommand{\hs}{\hspace{1mm}}

\newcommand{\apj}{ApJ}
\newcommand{\aap}{A\&A}
\newcommand{\apjl}{ApJL}
\newcommand{\mnras}{MNRAS}

\newcommand{\apjs}{ApJS}
\newcommand{\nat}{{\it Nature}}

\def\lsim{~\rlap{$<$}{\lower 1.0ex\hbox{$\sim$}}}

\def\gsim{~\rlap{$>$}{\lower 1.0ex\hbox{$\sim$}}}

\title[The Redshift Evolution of $f_{\rm esc}$ and the Ly$\alpha$ Fraction]{Evolution in the Escape Fraction of Ionizing Photons and the Decline in Strong Ly$\alpha$ Emission from $z>6$ Galaxies}

\author[Dijkstra et al]{Mark
Dijkstra$^{1,2}$\thanks{E-mail:mark.dijkstra@astro.uio.no}, Stuart Wyithe$^{3,4}$,  Zolt\'an Haiman$^5$, Andrei Mesinger$^6$, \newauthor \& Laura Pentericci$^7$
\\$^{1}$Max Planck Institute for Astrophysics, Karl-Schwarzschild-Str. 1, 85741, Garching, Germany\\$^2$Institute of Theoretical Astrophysics, University of Oslo, Postboks 1029, 0858 Oslo, Norway\\$^{3}$School of Physics,University of Melbourne, Parkville, Victoria, 3010, Australia\\$^{4}$ARC Centre of Excellence for All-Sky Astrophysics (CAASTRO)\\$^5$Department of Astronomy, Columbia University, 550 West 120th Street, New York, NY 10027, U.S.A.\\$^6$Scuola Normale Superiore, Piazza dei Cavalieri 7, 56126 Pisa, Italy\\$^7$ INAF - Osservatorio Astronomico di Roma, Via Frascati 33, IÐ00040, Monte Porzio Catone, Italy}

\begin{document}

\date{\today}
\pagerange{\pageref{firstpage}--\pageref{lastpage}} \pubyear{2012}

\voffset-.45in
\maketitle

\label{firstpage}
\begin{abstract}
The rapid decline in the number of strong Ly$\alpha$ emitting galaxies at $z>6$ provides evidence for neutral hydrogen in the IGM, but is difficult to explain with plausible models for reionization. We demonstrate that the observed reduction in Ly$\alpha$ flux from galaxies at $z>6$ can be explained by evolution in the escape fraction of ionizing photons, $f_{\rm esc}$. We find that the median observed drop in the fraction of galaxies showing strong Ly$\alpha$ emission, as well as the observed evolution of the Ly$\alpha$ luminosity function both follow from a small increase in $f_{\rm esc}$ of $\Delta f_{\rm esc} \sim 0.1$ from $f_{\rm esc}\sim0.6$ at $z\sim6$. This high escape fraction may be at odds with current constraints on the ionising photon escape fraction, which favor smaller values of $f_{\rm esc}\lsim20\%$. However, models that invoke a redshift evolution of $f_{\rm esc}$ that is consistent with these constraints can suppress the $z\sim7$ Ly$\alpha$ flux to the observed level, if they also include a small evolution in global neutral fraction of $\Delta x_{\rm HI}\sim0.2$. Thus, an evolving escape fraction of ionising photons can be a plausible part of the explanation for evolution in the Ly$\alpha$ emission of high redshift galaxies. More generally, our analysis also shows that the drop in the Ly$\alpha$ fraction is quantitatively consistent with the observed evolution in the Ly$\alpha$ luminosity functions of Ly$\alpha$ Emitters. 
\end{abstract}

 
\section{Introduction}
\label{sec:intro}

Observations of Ly$\alpha$ emitting galaxies are often interpreted to indicate that the intergalactic medium (IGM) is more opaque to Ly$\alpha$ photons at $z>6$, than at lower redshifts. In particular, while the Ly$\alpha$ luminosity function of Ly$\alpha$ selected galaxies (LAE) remains constant at $z=3-6$ \citep[e.g.][]{Hu98,Ouchi08}, it is observed to drop rapidly at $z>6$ \citep{Kashikawa06,Ouchi10,Clement}.  Since their rest-frame UV luminosity functions do not exhibit the same evolution \citep{Kashikawa06}, this reduction in the Ly$\alpha$ luminosity function at $z>6$ is likely caused by a reduction in the observed Ly$\alpha$ flux from these galaxies. Moreover, the so-called `Ly$\alpha$ fraction' -- which denotes the fraction of galaxies selected via the drop-out technique that exhibit strong Ly$\alpha$ emission lines -- increases between $z=3$ and $z=6$ \citep{Stark10,Stark11}, but then drops at $z>6$ \citep{Fontana10,Pentericci11,Schenker12,Ono12,Ca12,Treu12,Ca13,Treu13}. 

The observed reduction in Ly$\alpha$ flux from galaxies at $z>6$ is most readily explained by having additional intervening neutral atomic hydrogen, which is opaque to the Ly$\alpha$ flux, but not to the (non-ionizing) UV-continuum. This additional neutral atomic hydrogen is likely present naturally at $z>6$ when reionization has not been completed. Indeed, it has been predicted that the end of reionization should coincide with a reduction in the observed Ly$\alpha$ flux from galaxies \citep[e.g.][]{HS99}. 

Reionization was likely an inhomogeneous process in which fully ionized `bubbles' were seperated by neutral intergalactic gas \citep{F04,McK}. In this scenario, the progress of reionization is regulated by the growth of percolating HII bubbles. The final stages of reionization are characterized by the presence of large bubbles, whose individual sizes exceeded tens of cMpc \citep[e.g.][]{Zahn}. The majority of the galaxies we can detect with existing instruments preferentially resided inside large HII bubbles. Their Ly$\alpha$ photons would have been able to travel a significant distance before entering the neutral IGM, and redshift out of the Ly$\alpha$ resonance due to the Hubble expansion, which would facilitate their subsequent transmission through the neutral IGM \citep{M98,Cen05,McQuinn07, MF08LAE}.

Because of this effect it is difficult to interpret the observed reduction of Ly$\alpha$ flux from galaxies at $z>6$ in the context of inhomogeneous reionization models, since the imprint of the neutral IGM on the detectability of Ly$\alpha$ emission from galaxies is found to be subtle. For example, \citet{Dijk} showed that the observed drop in the Ly$\alpha$ fraction at $z>6$ requires a change in the volume averaged neutral fraction of hydrogen of $\Delta x_{\rm HI} \sim 0.5$ (also see Jensen et al. 2013, but see Taylor \& Lidz 2013 who caution that the required change can be reduced when effects of galaxy sample variance are accounted for), which is consistent with earlier constraints on this quantity by \citet{McQuinn07} and \citet{MF08LAE} from modeling the Ly$\alpha$ luminosity functions. As was pointed out by \citet{Dijk}, this required rapid evolution of $\Delta x_{\rm HI} \sim 0.5$ over such a short time is extreme\footnote{If the star formation rate is tied to the formation of dark matter structures (it is difficult to imagine that it can be faster than this), then it is limited by the growth of the collapsed mass. Even when one ignores recombinations in Lyman limit systems, which slow down the late stages of reionization (e.g. Furlanetto \& Mesinger 2008; Alvarez \& Abel 2012, Sobacchi \& Mesinger in prep), and when one assumes that the rapidly-growing exponential tail of the mass function is driving the late stages of reionization, then one still gets a $\lsim 50\%$ change in $x_{\rm HI}$ over $\Delta z =1$ (e.g. see Fig. 1 in Lidz et al. 2007). Relaxing both of these extreme assumptions slows down the evolution further.}. 

To alleviate this tension, \citet{Bolton13} recently showed that the required $\Delta x_{\rm HI}$ may be reduced significantly by Lyman limit systems (LLSs, self-shielding clouds with $N_{\rm HI}> 10^{17}$ cm$^{-2}$), whose number density may evolve rapidly at the end of reionization (although this rise is slowed down when recombinations in the IGM are taken into account, see Sobacchi \& Mesinger in prep). Similarly, the opacity of residual HI in the ionised IGM can increase by as much as $\sim 30\%$ between $z=5.7$ and $z=6.5$ \citep{Dijk07,Laursen11}. However, these models also predict that the IGM transmits low fraction of Ly$\alpha$ photons, $T_{\rm IGM}=0.1-0.3$. This is lower than current observational constraints on this quantity, which is likely related the impact of galactic winds on the Ly$\alpha$ line shape emerging from galaxies \citep{D13}. In any case, {\bf these possibilities} will need to be explored and quantified further in future work.

Ly$\alpha$ emission is powered by recombination following photoionization inside HII regions. The Ly$\alpha$ luminosity of a galaxy, $L_{\alpha}$, is therefore proportional to the total number of ionizing photons that do {\it not} escape from galaxies, i.e. $L_{\alpha} \propto (1-f_{\rm esc})\dot{N}_{\rm ion}$, where $f_{\rm esc}$ denotes the escape fraction of ionizing photons. Thus, having $f_{\rm esc}$ increase between $z=6$ and $z=7$ will reduce the Ly$\alpha$ luminosity of galaxies as observed. We expect this effect to be especially strong when $f_{\rm esc}$ is large. For example, the intrinsic Ly$\alpha$ luminosity $L_{\alpha}$ doubles when $f_{\rm esc}=0.8 \rightarrow 0.9$, while there is little impact when $f_{\rm esc}=0.1 \rightarrow 0.2$.

Modeling $f_{\rm esc}$ and its redshift dependence from first principles requires properly resolving the multi-phase structure of the interstellar medium \citep[see e.g.][]{FS11}, and the small spatial scales that are relevant for the transport of ionising radiation in high density gas \citep[e.g.][]{R13}. Direct observational constraints on the escape fraction, and it's redshift dependence are still highly uncertain. However, there are several lines of indirect evidence that $f_{\rm esc}$ increases with redshift. Measurements of the redshift dependence of the photoionization rate of the Ly$\alpha$ forest in combination with the observed redshift evolution of the UV-LF of drop-out galaxies, suggest that $f_{\rm esc}$ increases quite rapidly with redshift at $z \gsim 4$ \citep[e.g.][also see Inoue et al. 2006]{Kuhlen,Mitra13}. Moreover, the covering factor of low-ionization absorbers in drop-out galaxies has been observed to increase from $z=3$ to $z=4$, which provides independent evidence that $f_{\rm esc}$ is increasing with redshift \citep[][]{Jones12,Jones13}. This inferred redshift evolution could reflect e.g. ({\it i}) an evolution of the UV-emissivity per unit gas mass, ({\it ii}) an evolution in the clumpiness in the interstellar medium (ISM) of high-z galaxies which affects the covering factor of high-column density gas \citep{FS11}, and/or ({\it iii}) an evolution in the fraction of stars formed in low-mass halos which can efficiently `self-ionize' (e.g. Ferrara \& Loeb 2013).

The goal of this paper is to explore the impact of an increasing escape fraction on the visibility of the Ly$\alpha$ emission line from galaxies at $z>6$. We stress the purpose of our paper is a proof-of-concept, and that a systematic study will be performed in future work. The outline is as follows: we describe our model in \S~\ref{sec:model}, present our results in \S~\ref{sec:result}. We discuss our results in \S~\ref{sec:disc} and present our conclusions in \S~\ref{sec:conc}.

\section{The Model}
\label{sec:model}

\subsection{The EW-PDF and Its Redshift Evolution}
\label{sec:subsi}
The strength of the Ly$\alpha$ line emerging from galaxies is regulated by several physical processes: ({\it i}) The amount of Ly$\alpha$ that is produced, which depends on the initial mass function, stellar metallicity \citep[e.g.][]{Schaerer03}, and $f_{\rm esc}$; ({\it ii}) the amount of Ly$\alpha$ that escapes from the (dusty) interstellar medium of galaxies, which is correlated with the dust-content of galaxies \citep[][]{Atek09,Kornei10,Hayes11}; and ({\it iii}) the amount of Ly$\alpha$ that is scattered in the intergalactic/circum galactic medium. The Ly$\alpha$ line strength relative to the non-ionizing UV continuum is quantified by the (rest frame) equivalent width EW, i.e. EW$\equiv \frac{L_{\alpha}}{L_{\rm c}(\lambda_{\rm UV})}$, where $L_{\rm c}(\lambda_{\rm UV})$ denotes the luminosity density of the continuum at $\lambda_{\rm UV}=1300-1600$ \AA. The Ly$\alpha$ emitting properties of a collection of galaxies can be quantified by the (cumulative) EW-distribution function (PDF), denoted with $P({\rm EW})$. 

All processes listed above leave their imprint on the observed EW-PDF, and its redshift evolution. In particular, the EW-PDF evolves towards larger EWs from $z=3$ to $z=6$ \citep[][]{Stark10}, which can most easily be attributed to a decreasing dust content of galaxies in this redshift range \citep[][]{Hayes11}. Moreover, since the opacity of the intergalactic medium is expected to increase over the same redshift interval (and which would thus reduce the EW-PDF towards higher $z$), its impact appears subdominant to that of dust. However, the average\footnote{Of course, individual galaxies at $z>6$ can be dusty \citep[see e.g.][for an example at $z=7.5$ with weak Ly$\alpha$ emission]{Finkelstein13}.} dust content of galaxies keeps decreasing at $z>6$ \citep[e.g.][]{Finkcolors,Bouwenscolor}, and we would expect the EW-PDF to keep increasing. This is not what has been observed, and it becomes natural to search for alternative explanations for this sudden reduction in the EW-PDF at $z>6$. Suppression of the Ly$\alpha$ line by neutral intergalactic gas is a natural explanation, as there exist other observational indications for the presence of neutral intergalactic gas at $z>6$ \citep[e.g. in quasar absorption line spectra, see][]{WL04,MH07,Boltonqso}.

Previous models \citep[e.g.][]{Dijk,Jensen,Bolton13} have explored this possibility quantitatively, and assumed that the EW-PDF at $z=7$ only differs from the one measured at $z=6$ due to the intervening neutral IGM. Under this assumption we have EW$_7$=EW$_6\times \mathcal{T}_{\rm IGM,7}/\mathcal{T}_{\rm IGM,6}$, where $\mathcal{T}_{\rm IGM,x}$ corresponds to the IGM transmission fraction at redshift $z=x$. We can then derive the EW-PDF at $z=7$ from the one observed at $z=6$ simply from\footnote{\citet{Dijk} computed $\mathcal{T}_{\rm IGM}$-PDFs, $P_7(\mathcal{T}_{\rm IGM})$, as part of their analysis, and this could also be folded into the calculation. \citet{Dijk} further assumed that $\mathcal{T}_{\rm IGM,6}=1$. Under these assumptions we have $P_7({\rm EW}) \propto \int d\mathcal{T}_{\rm IGM}\hs P_7(\mathcal{T}_{\rm IGM})P_6({\rm EW}/\mathcal{T}_{\rm IGM})$.} $P_7({\rm EW}) \propto P_6({\rm EW}\times \mathcal{T}_{\rm IGM,6}/\mathcal{T}_{\rm IGM,7})$. Because it is difficult to explain the observed $z-$evolution with reionization alone (see \S~\ref{sec:intro}), we perform a similar analysis which includes a redshift escape fraction in this paper, as is discussed in more detail next.

\subsection{The Redshift Evolution of the EW-PDF with $f_{\rm esc}(z)$}

To incorporate the impact of $f_{\rm esc}$, we first introduce the distribution $P_0({\rm EW})$ which denotes the EW-PDF, if the escape fraction of ionizing photons at $z=6$ were $f_{\rm esc}=0$. This model maximizes the produced Ly$\alpha$ luminosity/EW for a given IMF and stellar metallicity. We assume that $P_0({\rm EW})$ is an exponential function at EW$>0$ with a scalelength EW$_0$, zero otherwise. We next assume that galaxies have a distribution of $f_{\rm esc}$, which we denote with\footnote{We denote differential and cumulative probability distributions with lower and upper case letters, respectively. That is, we denote $p(x) \equiv \frac{dP}{dx}$.} $p(f_{\rm esc})$. In this case we have
\begin{equation}
 P_x({\rm EW}) \propto \int P_0({\rm EW}/[1-f_{\rm esc}])p_{\rm x}(f_{\rm esc})df_{\rm esc},
 \label{eq:eq1}
 \end{equation} where $p_{\rm x}(f_{\rm esc})$ denotes the differential distribution of $f_{\rm esc}$, and where $P_{\rm x}({\rm EW})$ denotes the cumulative distribution of EW at redshift $z=x$. We assume that $p_{\rm x}(f_{\rm esc})$ is Gaussian when $0 \leq f_{\rm esc} \leq 1$, and zero otherwise. We denote the mean, or expectation value, of $f_{\rm esc}$ with $\langle f_{\rm esc} \rangle \equiv \int_0^1 df_{\rm esc} f_{\rm esc} p(f_{\rm esc})$. For a given $p_{\rm x}(f_{\rm esc})$, we constrain the scalelength EW$_0$ by comparing to the observed $P_{x}({\rm EW})$. We point out that the expression for $P_x({\rm EW})$ in Eq~\ref{eq:eq1} is a weighted sum of exponential functions with different scale-lengths. The final function is therefore generally not an exponential function, which differs from previous analyses \citep[as in][]{Dijk,Jensen,Bolton13}.  
In our analysis we make the following additional assumptions: 
 \begin{itemize}

\item  $f_{\rm esc}$ at a fixed $z$ is the same for all galaxies: i.e. $f_{\rm esc}$ does not depend on $M_{\rm UV}$. As discussed in \S~\ref{sec:fescmuv}, this assumption is likely unrealistic. However, the precise $M_{\rm UV}$-dependence of $f_{\rm esc}$ is still poorly constrained. Instead of including this possible dependence in our models, we opt to study simpler models, which can provide a baseline for more complex future models that include this effect. Moreover, the range of UV-luminosities of the galaxies we are modeling is still limited, and the impact of $f_{\rm esc}$ depending on $M_{\rm UV}$ does not affect our main results. 

\item First, we assume that the EW-PDF at $z=7$ differs only from the one at $z=6$ due to the a change in the escape fraction of ionizing photons, $f_{\rm esc}$ (see \S~\ref{sec:subsi} for a motivation of this assumption). We present our results on this analysis in \S~\ref{sec:result}. We discuss how our analysis is modified if we assume that the EW-PDF evolves at $z>6$ as a result of a joint evolution in $f_{\rm esc}(z)$ and reionization in \S~\ref{sec:addreion}. We generally present our results in terms of the expectation values $\langle f_{\rm esc,7}\rangle \equiv \langle f_{\rm esc,6}\rangle +\Delta \langle f_{\rm esc} \rangle \leq1$.  We point out that the large impact on the redshift evolution in $P({\rm EW})$ for large $\langle f_{\rm esc,6}\rangle$ at fixed $\Delta \langle f_{\rm esc} \rangle$ (as mentioned in \S~\ref{sec:intro}) arises because $f_{\rm esc}$ appears as $(1-f_{\rm esc})^{-1}$ in the argument of $P_0({\rm EW})$ in Eq~\ref{eq:eq1}, which increases rapidly when $f_{\rm esc} \rightarrow 1$. 

\item Our analysis implicitly assumes that the $z-$evolution of $f_{\rm esc}$ does not affect the $z-$evolution of other quantities: e.g. increasing the escape fraction does not increase ({\it i}) the fraction of Ly$\alpha$ photons that are transmitted through the IGM, or ({\it ii}) the fraction of Ly$\alpha$ photons that can escape from the dusty ISM of galaxies. As we argue in \S~\ref{sec:fescvsz}, ({\it i}) is reasonable, while ({\it ii}) presents a limitation of the model, but may help explain why the redshift evolution of the escape fraction may only have noticeable effects on the visibility of the Ly$\alpha$ line at $z>6$.
 \end{itemize}

\subsection{The Redshift Evolution of the EW-PDF with $f_{\rm esc}(z)$ and $x_{\rm HI}(z)$}
\label{sec:addreion}
\begin{figure*}
\vbox{\centerline{\epsfig{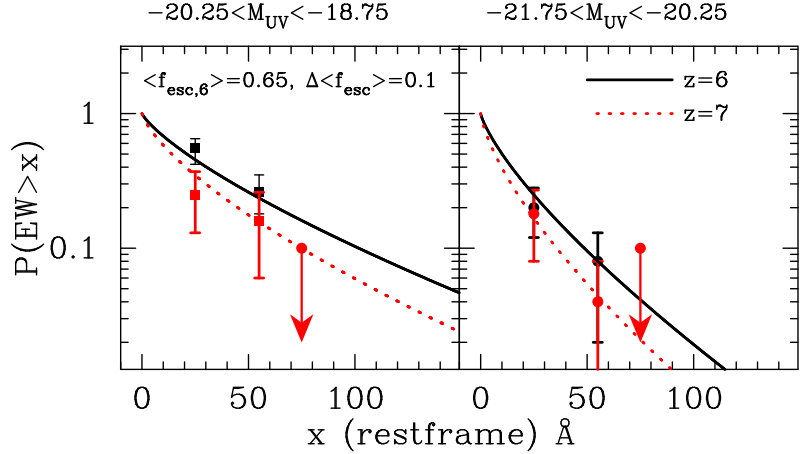}}}
\caption[]{This Figure shows cumulative Ly$\alpha$ EW distributions for a sample of UV-faint ({\it left panel}) and UV-bright ({\it right panel}) drop-out selected galaxies. {\it Black filled squares}/{\it red filled circles} represent observations at $z\sim6$ /$z\sim 7$ (taken from Ono et al. 2012). The {\it black solid line} represents a models at $z=6$ assuming ({\it i}) that $p_6(f_{\rm esc})$ is a Gaussian with $\sigma=0.3$ and $\langle f_{\rm esc} \rangle =0.65$, and that ({\it ii}) $P_0({\rm EW}) \propto \exp[-{\rm EW}/{\rm EW}_0]$. Here we took EW$_0=110$ \AA/EW$_0=55$ \AA\hs to match the UV-faint/UV-bright data. {\it Red dotted lines} represent predictions if we only modify $P_7({\rm EW})$ such that $\Delta \langle f_{\rm esc} \rangle=0.1$ (i.e. $\langle f_{\rm esc} \rangle =0.75$). This Figure illustrates that it is possible to explain the observed drop in Ly$\alpha$ fraction completely with a modest increase in $f_{\rm esc}$, the escape fraction of ionizing photons, provided that $f_{\rm esc}$ for LAE is already high at $z\approx6$.}
\label{fig:fig1}
\end{figure*}
If we add the effect of an (modest) evolution in reionization together with a modest evolution in $f_{\rm esc}$, then we can achieve a larger drop in the Ly$\alpha$ fraction.  The additional impact of reionization on the EW-PDF can be incorporated as\footnote{This (again) assumes that $\mathcal{T}_{\rm IGM}(z=6)=1$. This assumption is conservative in the sense that it maximizes the impact of reionization \citep[see][for a discussion]{Dijk}.}
\begin{eqnarray}
 P_x({\rm EW}) \propto \int d\mathcal{T}_{\rm IGM}\hs p_{\rm x}(\mathcal{T}_{\rm IGM})\times \\ \nonumber \times \int P_0({\rm EW}/[1-f_{\rm esc}]\mathcal{T}_{\rm IGM})p_{\rm x}(f_{\rm esc})df_{\rm esc},
 \label{eq:eq1}
 \end{eqnarray} where $p(\mathcal{T}_{\rm IGM})$ denotes the $\mathcal{T}_{\rm IGM}$-PDF. Computing this PDF was the prime focus of the analysis described in \citet[][]{Dijk}, who combined galactic outflow models with large-scale semi-numeric simulations of reionization. 
 
Following \citet{Dijk}, we model the impact of the galactic outflow on the Ly$\alpha$ photons emerging from the galaxy using spherical shell models. For the analysis in this paper, we used shells with $N_{\rm HI}=10^{20}$ cm$^{-2}$ and wind velocities of $v_{\rm wind}=25$ km s$^{-1}$. For more a detailed discussion on these models the reader is referred to \citet{DW10} and \citet{Dijk}. We simulate inhomogeneous reionization at $z=7$ with the publicly-available, semi-numerical code DexM (\citealt{MF07}; http://homepage.sns.it/mesinger/Sim).  DexM combines excursion set and perturbation formalisms to generate various cosmic fields, and has been extensively tested against numerical simulations \citep{MF07, MFC11, Zahn}. Our simulation box is 200 Mpc on a side with a resolution of 500$^3$.  We resolve halos down to a minimum mass of $M_{min}\gsim5\times10^8 M_\odot$, consistent with the expected cooling threshold at $z\sim7$ \citep{SM13b}.  The simulations present minor modifications of those used in Dijkstra et al. 2011 (specifically the simulated redshift has been changed for the problem at hand, and the minimum mass has been increased by a factor of $\sim 5$ to account for photoionization feedback). These modified simulations will be described in detail in a subsequent work, Mesinger et al., in prep. We note that the new $p(\mathcal{T}_{\rm IGM})$ look very similar to those we obtained with the original simulations that were used in \citet{Dijk}.

\section{Results}
\label{sec:result}

\subsection{Evolution in the Ly$\alpha$ Fraction from Redshift Evolution in $f_{\rm esc}(z)$}
Figure~\ref{fig:fig1} shows cumulative Ly$\alpha$ EW-PDFs of drop-out galaxies with $-20.25< M_{\rm UV}<-18.75$ ({\it left panel}) and $-21.75< M_{\rm UV}<-20.25$ ({\it right panel}). The {\it black squares} ({\it red filled circles}) represent data points at $z=6$ ($z=7$) from the compilation by \citet{Ono12}. This Figure also shows an example of a model in which the observed redshift evolution of the Ly$\alpha$ EW-PDF can be mimicked completely with a redshift-dependent $f_{\rm esc}(z)$.

The {\it black solid line} in the {\it left panel} shows a model EW-PDF at $z=6$ which assumes ({\it i}) that $p_6(f_{\rm esc})$ is Gaussian with a standard deviation $\sigma=0.3$ and $\langle f_{\rm esc} \rangle =0.65$, and that ({\it ii}) $P_0({\rm EW}) \propto \exp[-{\rm EW}/{\rm EW}_0]$, where we took EW$_0=110$ \AA\hs to match the $z=6$ data. These model parameters were chosen to quantitatively illustrate our main point\footnote{We stress that: ({\it i}) the choice $\sigma=0.3$ is a bit arbitrary, but intermediate between having negligible dispersion and having such a large dispersion that $p(f_{\rm fesc})$ approaches a uniform distribution. We have verified that our main conclusions are not affected by the precise choice of $\sigma$; ({\it ii}) EW$_0$ corresponds to the scale-length of $P_0({\rm EW})$. As we mentioned in \S~\ref{sec:model}, we use Eq~1 to compute $P_6({\rm EW})$. In contrast, previous works adopted an exponential function for $P_6({\rm EW})$, which generally has a different scale-length than $P_0({\rm EW})$.}. The {\it black solid line} shown in the {\it right panel} represents the same model, but with EW$_0=55$ \AA. This reduced value\footnote{Coincidentally the scalelength EW$_0=55$ \AA\hs in $P_0({\rm EW})$ for the sample of bright drop-out galaxies is close to the scale-length of EW$_{\rm 0}=50$ \AA\hs adopted in Dijkstra et al. (2011) for the function $P_6({\rm EW})$, the observed EW-distribution of $z\sim 6$ drop-out galaxies (from Stark et al. 2010).} of EW$_0$ reflects that the Ly$\alpha$ fraction decreases towards brighter $M_{\rm UV}$. 

The {\it red dotted lines} represent a model in which we only modified $P_7({\rm EW})$ such that $\langle f_{\rm esc,7}\rangle=0.75$, i.e. $\Delta \langle f_{\rm esc} \rangle=0.1$. This model can fully explain the observed reduction of the Ly$\alpha$ EW-PDF and the drop in the Ly$\alpha$ fraction. 

\subsection{Evolution in the Luminosity Functions from Redshift Evolution in $f_{\rm esc}(z)$}

We also show the impact of a changing $f_{\rm esc}$ on the Ly$\alpha$ luminosity function of LAEs. For this exercise, we follow the procedure of \citet{DW12}, who constructed Ly$\alpha$ luminosity functions by combining observed UV luminosity functions of drop-out selected galaxies with observed Ly$\alpha$ equivalent width distributions. \citet{DW12} found that this procedure reproduces observed luminosity functions of LAEs well following inclusion of a rescaling by a factor of $F=0.5$ (for a detailed discussion we refer the interested reader Dijkstra \& Wyithe 2012). Importantly, the redshift evolution of the luminosity functions of LAEs was reproduced well at all redshifts $z\gsim 3$, which is important for the analysis presented here. 

The differential Ly$\alpha$ luminosity function, denoted by $\frac{dn}{d\log L_{\alpha}}d\log L_{\alpha}$, measures the comoving number density of galaxies with (the logarithm of their) Ly$\alpha$ luminosities in the range $\log L_{\alpha} \pm d\log L_{\alpha}/2$, and is given by
\begin{eqnarray}
\nonumber
\frac{dn}{d\log L_{\alpha}}=F\int_{M_{\rm min}}^{M_{\rm max}}dM_{\rm uv}\textcolor{black}{\phi(M_{\rm UV},z)}\textcolor{black}{\frac{dP}{d \log L_{\alpha}}(M_{\rm uv},z)}.
\label{eq:phi}
\end{eqnarray} Here, $\phi(M_{\rm UV})dM_{\rm UV}$ denotes the comoving number density of drop-out selected galaxies with absolute magnitudes in the range $M_{\rm UV} \pm dM_{\rm UV}/2$, for which we adopted the Schechter function parameterization given by Table~1 in \citet{Bouwens12}. The distribution of Ly$\alpha$ luminosity is $\frac{dP}{d \log L_{\alpha}}(M_{\rm uv})$$=\ln 10\times{\rm EW}\times \frac{dP}{d{\rm EW}}(M_{\rm UV})$. Finally, we have $F=0.5$ (as mentioned above). Equation~\ref{eq:phi} therefore allows us to compute the redshift evolution in the Ly$\alpha$ luminosity function due to both evolution in the UV-luminosity function of drop-out galaxies, and the redshift evolution of $\frac{dP}{d{\rm EW}}$ (i.e. the Ly$\alpha$ fraction).

The {\it black solid line} shows the predicted Ly$\alpha$ luminosity function at $z=5.7$, where we assumed that $\frac{dP}{d{\rm EW}}(z=5.7)=\frac{dP}{d{\rm EW}}(z=6.0)$, and we adopted the $\frac{dP}{d{\rm EW}}(z=6.0)$ shown in the {\it left panel} of Figure~\ref{fig:fig1} (i.e. the EW-PDF that described the UV-faint population of drop-out galaxies)\footnote{The majority of $z\sim 5.7$ LAEs have $M_{\rm UV}\gsim -20.5$, where LAEs are detected only marginally in the continuum \citep[see Fig~22 of][]{Ouchi08}. A proper calculation would take into account that there are UV bright LAEs ($M_{\rm UV} \lsim -21.0$) for which $P({\rm EW})$ is different. However, because there are fewer of these galaxies, the impact of the assumed $P({\rm EW})$ at bright $M_{\rm UV}$ only has little impact on the bright end of the Ly$\alpha$ luminosity function. We have repeated our analysis with the $M_{\rm UV}$-dependent $P({\rm EW})$ given by Dijkstra \& Wyithe (2012) and obtained practically the same results.}. 
The model luminosity function fits the observations of \citet{Ouchi08} ({\it blue filled circles}) well. The {\it red lines} show the predicted luminosity functions at $z=6.5$. The {\it dashed red line} assumes that the EW-PDF at $z=6.5$ is the same as at $z=6$, while the {\it dotted red line} assumes that $f_{\rm esc}$ evolved from $\langle f_{\rm esc}\rangle=0.65$ at $z=6$ to $\langle f_{\rm esc} \rangle=0.70$ at $z=6.5$ (i.e. this corresponds to the model shown in Fig~\ref{fig:fig1} and we assumed that $\frac{d\langle f_{\rm esc}\rangle}{dz}=0.1$ and $\langle f_{\rm esc,6}\rangle=0.65$). The data at $z=6.5$ ({\it red filled squares}, taken from Ouchi et al. 2010) does not favor any of the models significantly. The {\it blue lines} show the predicted luminosity functions at $z=7.0$, where {\it the dotted line} shows a model in which the $f_{\rm esc}$ evolved to $\langle f_{\rm esc,7} \rangle =0.75$ at $z=7.0$. This model lies much closer to the observations of Ota et al. (2010, {\it green diamonds}) than the model which keeps $f_{\rm esc}$ constant. Thus, evolution in the escape fraction of ionizing photons simultaneously explains the observed drop in the Ly$\alpha$ fraction between $z=6$ and $z=7$, and the observed evolution in the Ly$\alpha$ luminosity functions within the same redshift range. More generally, our analysis shows that the drop in the Ly$\alpha$ fraction is quantitatively consistent with the observed evolution in the Ly$\alpha$ luminosity functions of LAEs. 
\begin{figure}
\vbox{\centerline{\epsfig{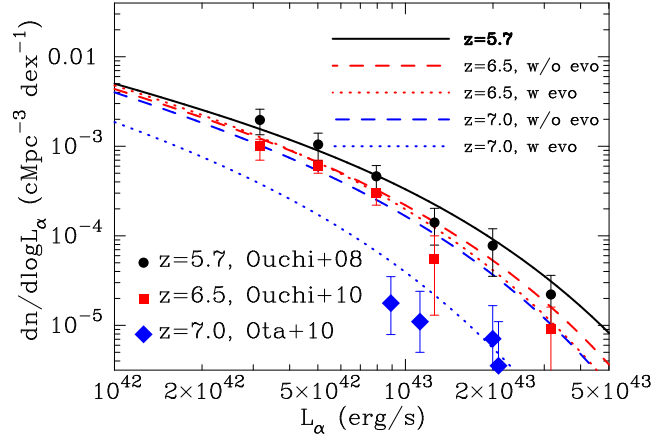}}}
\caption[]{This Figure compares predicted Ly$\alpha$ luminosity functions to observations at $z=5.7$ ({\it blue filled circles}, taken from Ouchi et al. 2008), at $6.5$ ({\it red filled squares}, taken from Ouchi et al. 2010), and $z=7.0$ ({\it green diamonds}, taken from Ota et al. 2010). The {\it black solid line} represents our model at $z=5.7$ which uses the z=6 EW-PDF shown as the {\it black solid} in the {\it left panel} of Figure~\ref{fig:fig1}. The {\it blue} and {\it red dotted lines} represent predicted luminosity functions at $z=6.5$ and $z=7.0$ when we assume the same redshift evolution in $f_{\rm esc}$ as in Figure~\ref{fig:fig1}. The {\it blue} and {\it red dashed lines} assume no evolution in the Ly$\alpha$ EW-PDFs (see text). This Figure shows that the model that reproduced the observed drop in the Ly$\alpha$ fraction (shown in Fig~\ref{fig:fig1}), also naturally reproduces the observed evolution in the Ly$\alpha$ luminosity functions of LAEs.}
\label{fig:fig3}
\end{figure}

\subsection{Evolution in the Ly$\alpha$ Fraction from Joint Redshift Evolution in $f_{\rm esc}(z)$ and $x_{\rm HI}$}
\label{sec:fescz}

The previous section illustrated that a small evolution in $f_{\rm esc}$ from a value of $\langle f_{\rm esc,6}\rangle=0.65$ is sufficient to explain the observed decrease in the LAE luminosity function and the fraction of strong Ly$\alpha$ emitting galaxies. However this high escape fraction may be at odds with constraints obtained from measurements of Thomson optical depth to the cosmic microwave background, the photoionization rate of the Ly$\alpha$ forest, and the observed redshift evolution of the UV-LF of drop-out galaxies, which jointly favour values of $f_{\rm esc,6}\sim10-20\%$ (Inoue et al. 2006, Wyithe et al. 2010, Kuhlen \& Faucher-Gigu{\`e}re 2012, Robertson et al. 2013, Becker \& Bolton 2013, also see Finkelstein et al. 2012). While these estimates rely on uncertain extrapolations\footnote{Figure~8 of  Kuhlen \& Faucher-Gigu{\`e}re (2012) shows the constraint on $f_{\rm esc}(z=7)$ as a function of the minimum UV-luminosity to which they extrapolate the observed UV-LFs. Notably, the constraint $f_{\rm esc}(z=7) \sim 20\%$ includes all galaxies with $M_{\rm UV} < -13.0$ and thus involves extrapolating the UV-LF by 4-5 magnitudes.} of the luminosity density in low luminosity galaxies (and towards higher redshifts), the first direct constraints on $f_{\rm esc}$ in LAEs have recently been reported by \citet{Ono10} who found $f_{\rm esc,6} \lsim 0.6$ ($1-\sigma$).

Thus, the escape fraction evolution is unlikely to fully explain the observed drop in the Ly$\alpha$ fractions. We therefore examine the specific case in which $f_{\rm esc}$ evolves as $f_{\rm esc}(z)=f_{\rm 0}([1+z]/5)^{\kappa}$ \citep[as in][also see Becker \& Bolton 2013]{Kuhlen}, and adopt $\kappa=4$ and $f_{\rm 0}=0.04$ \citep[which is consistent with observations, see][]{Kuhlen}\footnote{The evidence for such a dramatic evolution in $f_{\rm esc}$ can be ameliorated however with, e.g.: (i) a luminosity (i.e. halo mass) dependence of $f_{\rm esc}$ (as in e.g. Ferrara \& Loeb 2013); (ii) an (evolving) contribution of very faint dwarf galaxies; and/or (iii) a loosening of the kinetic Sunyaev-Zeldovich  reionization constraints used in the analysis for \citet{Kuhlen}, which \citet{MMS12} show are insensitive to the end stages of reionization.  We return to the issue of (i) in the next section, but (ii) in particular is not surprising, as pointed out by \citet{Alvarez12}.  A population of dwarf galaxies near the atomic cooling threshold with moderate values of $f_{\rm esc}$ could reionize the Universe at $z\sim10$.  Photo-ionization feedback from reionization itself would subsequently suppress star formation in galaxies around this mass scale (e.g. \citealt{SM13b}), so that they no longer contribute to the ionizing emissivity at $z \sim 4$.}. For $f_{\rm 0}=0.04$, we have $f_{\rm esc}=0.15$ at $z=6$ and $f_{\rm esc}=0.26$ at $z=7$.
While this model predicts that $\Delta \langle f_{\rm esc} \rangle \sim 0.1$, it does not suffice to fully explain the observed evolution of the EW-PDF.  
This is due to the smaller value of $\langle f_{\rm esc,6} \rangle$ than what we assumed previously, which gives rise to a weaker impact on the redshift evolution of the EW-PDF (see \S~\ref{sec:intro}, and \S~\ref{sec:model}).

\begin{figure}
\vbox{\centerline{\epsfig{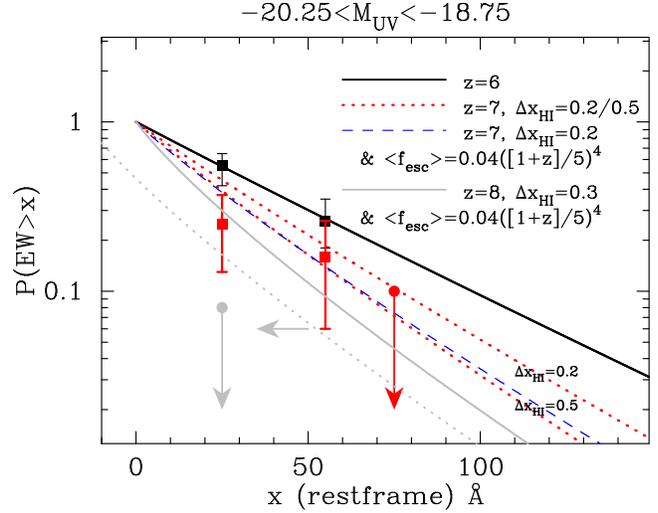}}}
\caption[]{This Figure shows the cumulative Ly$\alpha$ EW distribution for UV-faint drop-out selected galaxies (see Fig~\ref{fig:fig1} for a description of the lines and data points, and the text for details on the models). Here, the upper/lower {\it red dotted line} represents a model in which we modify the EW-PDF by having the IGM opacity increase due to an increase in the globally averaged neutral fraction, $\Delta x_{\rm HI}=0.17$/$\Delta x_{\rm HI}=0.5$. The {\it dashed blue lines} represents a model in which in addition $f_{\rm esc}$ evolves as $f_{\rm esc}(z)=0.04([1+z]/5)^{4}$ in addition to having $\Delta x_{\rm HI}=0.17$. This Figure shows that mild evolution in both $x_{\rm HI}$ and $f_{\rm esc}$ can mimick a more rapid evolution in $x_{\rm HI}$ and thus explain the observed drop in Ly$\alpha$ fractions. The {\it grey solid lines} show predictions if we extrapolated the redshift evolution of $f_{\rm esc}$ to $z=8$, while also changing the globally averaged neutral fraction to $x_{\rm HI}=0.3$. This prediction is still at odds with recently inferred fraction at $z=8$ by \citet{Treu13} (represented by the upper limit at EW$=25$ \AA). The models can be made more consistent with this upper limit if we shift the predicted EW-PDFs by $\Delta$EW=-25 \AA\hs (shown by the {\it grey dotted line}, see text).}
\label{fig:fig4}
\end{figure}

In Figure~\ref{fig:fig4} we show the cumulative Ly$\alpha$ EW distribution for UV-faint drop-out selected galaxies only (see Fig~\ref{fig:fig1} for a description of the lines and data points; the following results are quantitatively the same for the UV-bright galaxies). The {\it black solid line} now represents a model in which ({\it i}) $P_6(f_{\rm esc})$ is a Gaussian with $\langle f_{\rm esc,6} \rangle =0.15$ and $\sigma=0.3$, and ({\it ii}) $P_0({\rm EW}) \propto \exp[-{\rm EW}/{\rm EW}_0]$, where EW$_0=55$ \AA\hs (note that it was\footnote{We have to pick a different $EW_0$ because we changed $p_6(f_{\rm esc})$. For the new choice for $p_6(f_{\rm esc})$ the predicted EW-PDF $P_6(\rm EW)$ differs from that shown in Fig~1.} EW$_0=110$ \AA\hs in the model studied in \S~\ref{sec:result} for the UV-faint sample). The {\it red dotted lines} show the predicted changes in the EW-PDF under the assumption that the drop is entirely due to a changing ionization state of the IGM (as in Dijkstra et al. 2011 and Jensen et al. 2013)\footnote{It is worth noting that if the IGM is not fully neutral even at $z\approx6$ (which is consistent with current observations (e.g. \citealt{MMF11, SMH13}), then the $\Delta x_{\rm HI}$ required to match observations is decreased \citep{McQuinn07, MF08LAE, Dijk} }. The upper/lower {\it red dotted line} corresponds to $\Delta x_{\rm HI}=0.2$/$\Delta x_{\rm HI}=0.5$. The {\it blue dashed line} represents a model in which $\Delta  \langle f_{\rm esc} \rangle =0.11$ in addition to having $\Delta x_{\rm HI}=0.2$. {\it This model is virtually indistinguishable from the model with $\Delta x_{\rm HI}=0.5$ and no evolution in $f_{\rm esc}$.} Thus, extrapolation of the observed evolution in $f_{\rm esc}$ is equivalent to having an additional $\Delta x_{\rm HI}=0.3$ between $z=6$ and $z=7$ in effecting the properties of Ly$\alpha$ flux and EW. 
 
The {\it grey solid line} represents the a model in which the redshift evolution in $f_{\rm esc}$ is extrapolated to $z=8$. In this model, we additionally assume that the globally averaged neutral fraction, $x_{\rm HI}$, has evolved further to $x_{\rm HI}=0.3$. Figure~\ref{fig:fig4} shows that an evolving $f_{\rm esc}(z)$ has a dramatic impact on the predicted Ly$\alpha$ fraction at $z=8$. However, even these models do not reproduce the recently inferred Ly$\alpha$ fraction at $z=8$ by \citet{Treu13} (represented by the upper limit at EW$=25$ \AA). This `failure' can be partially remedied by requiring that $x_{\rm HI} \gg 0.3$ at $z=8$, which would again require a very rapid evolution in $x_{\rm HI}$. It may also be related to the fact that our models assume that all drop-out galaxies have a Ly$\alpha$ emission line with EW$>0$, i.e. that $P_0({\rm EW}>0)=1$. While this assumption is consistent with observations of drop-out galaxies at $z=6$, the observational uncertainties allow us to relax this assumption and shift the intrinsic distribution by, say, $\Delta$EW=$-25$ \AA, which would imply that $P_0({\rm EW}>0)<1$ as observed in $z\sim 3$ drop-out galaxies \citep{Shapley03}. While the escape fraction of Ly$\alpha$ photons increases with redshift, it is unclear whether $P_0({\rm EW}>0)=1$ at $z \geq 6$. If we apply a shift of $\Delta$EW=$-25$ \AA\hs to the predictions at $z=8$, then our model predictions lie much closer to the upper limit at $z=8$ (as shown by the {\it grey dotted line}). \\

\section{Discussion}
\label{sec:disc}

\subsection{A Mass/Luminosity-dependent $f_{\rm esc}(z)$?}
\label{sec:fescmuv}

A caveat is that the observationally inferred redshift evolution of $f_{\rm esc}$ adopted in \S~\ref{sec:fescz}, refers to an average over the entire galaxy population. It has been argued that the inferred redshift evolution may be driven by a mass and/or luminosity dependence of the escape fraction \citep{Alvarez12,FL13}.  In this scenario, $f_{\rm esc}$ decreases towards higher masses and/or luminosities, and the population averaged escape fraction reflects the redshift evolution of the luminosity and/or mass functions. Support for the mass and/or luminosity dependence of the escape fraction is provided by the observationally inferred escape fraction of LAEs at $z=3$ of $f_{\rm esc}\sim 0.1-0.3$, which is significantly higher than the inferred fraction for the more massive LBGs where $f_{\rm esc}\sim 0.05$ \citep[e.g.][]{Iwata09,Nestor11}. 

This suggests that assuming a uniform $f_{\rm esc}(z)$ is not realistic. However, the higher inferred escape fraction of LAEs at $z=3$ of $f_{\rm esc}(z=3)\sim 0.1-0.3$ may imply that we need a less rapid evolution in $f_{\rm esc}(z)$ to have a significant impact on the redshift evolution of the Ly$\alpha$ fraction: we showed that $\Delta \langle f_{\rm esc} \rangle =0.1$ can either help explain the observed drop in the Ly$\alpha$ fraction between $z=6$ and $z=7$ if $\langle f_{\rm esc,6} \rangle =0.15$, or explain the complete evolution when $\langle f_{\rm esc,6} \rangle =0.15$. If $f_{\rm esc}(z)$ increases continuously between $z=3$ and $z=6$, then $\langle f_{\rm esc,6} \rangle$ may be large enough that only a small additional change may have a significant impact. As mentioned in \S~\ref{sec:intro}, the decreasing covering factor of low ionization absorption lines with redshift does suggest that escape fraction increases with redshift for a fixed galaxy population \citep[][]{Jones12,Jones13}.

\subsection{Ly$\alpha$ Transport \& Escape, and a Bimodal $p(f_{\rm esc})$}
The observed Ly$\alpha$ flux of a galaxy not only depends on $f_{\rm esc}$, but also on the effective escape fraction of Ly$\alpha$ photons from the galaxies to the observer, $f^{\rm eff}_{\rm esc,Ly}$. Thus we have $L_{\alpha} \propto f^{\rm eff}_{\rm esc,Ly}(1-f_{\rm esc})$. The effective escape fraction $f^{\rm eff}_{\rm esc,Ly}$ includes the fraction of Ly$\alpha$ photons that escape from the interstellar medium of galaxies, but also the fraction that is subsequently transmitted through the intergalactic medium \citep[IGM, e.g.][]{D13}. Both of these processes may depend on the value of $f_{\rm esc}$, i.e. $ f^{\rm eff}_{\rm esc,Ly}= f^{\rm eff}_{\rm esc,Ly}(f_{\rm esc})$ (see discussion in \S~\ref{sec:fescvsz}).

The escape of Ly$\alpha$ photons from a dusty interstellar medium for example, is a complex process which depends on the dust content of the ISM, as well as its kinematics \citep[e.g.][]{Atek08,Hayes11}. Irrespective of these complexities, the escape of ionizing photons requires low column density ($N_{\rm HI} < 10^{17}$ cm$^{-2}$) sightlines out of the galaxy. If these low HI-column density paths are surrounded by higher column density sightlines which are opaque to ionizing photons \citep[as in the `blow-out' model proposed by][]{Nestor11}, then we may expect Ly$\alpha$ photons to scatter and preferentially escape along these same paths (see Behrens et al. 2014 for a more detailed investigation of this effect). This can introduce a correlation between the escape fractions of Ly$\alpha$ and ionizing photons. In the most extreme case, individual galaxies would have a bimodal distribution for $p(f_{\rm esc})$ which contains peaks at $f_{\rm esc}=0$ and $f_{\rm esc}=1$. Interestingly, there is observational support for such a bimodality in observations of star forming galaxies, which indicate that a small fraction has a large $f_{\rm esc}$, while a large fraction practically has $f_{\rm esc}=0$ \citep[][]{Shapley06,Nestor11,Vanzella12}. However, even in this scenario it is the escape fraction averaged\footnote{In the extreme scenario in which $f_{\rm esc}=1$ along certain sightlines, and $f_{\rm esc}=0$ otherwise,  the angle-averaged escape fraction $f^{\Omega}_{\rm esc}$ is just the sky-covering factor of low column density `holes', i.e $f^{\Omega}_{\rm esc}=\Omega_{\rm hole}/4 \pi$.} over all sightlines, $f^{\Omega}_{\rm esc}$, that is relevant for powering nebular emission (and also for reionizing the Universe). The models described in this paper therefore also describe a scenario in which $f^{\Omega}_{\rm esc}$ is distributed as a Gaussian. \\

It is nevertheless good to keep in mind that our scenario does not describe the more extreme situation in which $f^{\Omega}_{\rm esc}$ has a bimodal distribution, i.e. it does not describe a scenario in which some galaxies have $f_{\rm esc}\gg 0$  while others have $f_{\rm esc}=0$ {\it in all directions}. In this scenario -- which appears to be at odds with observed covering factors of low-ionization absorption line systems in drop-out galaxies, which are typically $<1$ \citep[see e.g.][]{Heckman11,Jones13}-- an increase in $f_{\rm esc}(z)$ with $z$ translates to an increase in the fraction of galaxies with $f_{\rm esc} \gg 0$. This would also reduce the fraction of star forming galaxies that produce Ly$\alpha$ photons. However, the overall reduction in the number density of LAEs would be weaker than in our models.

\subsection{Dependence of $f^{\rm eff}_{\rm esc,Ly}$ on $f_{\rm esc}$}
\label{sec:fescvsz}

The discussion above shows that the effective escape fraction of Ly$\alpha$ photons can depend on $f_{\rm esc}$ (and therefore possibly on $M_{\rm UV}$ as in Forero-Romero et al. 2012). If ionizing photons escape anisotropically, then so do Ly$\alpha$ photons. However, the first calculations of this effect have been reported only recently (Behrens et al. 2014). The precise correlation this may introduce between $f_{\rm esc}$ and $f^{\rm eff}_{\rm esc,Ly}$ is complex (it depends e.g. on the geometry of the low-column density holes, outflow properties), and has therefore not been quantified yet. If we assume for simplicity that $f^{\rm eff}_{\rm esc,Ly} \propto f_{\rm esc}^y$, then we expect the observed Ly$\alpha$ luminosity to increase with $f_{\rm esc}$ until $f_{\rm esc,pk}=\frac{y}{y+1}$, after which it decreases.\\

The correlation between $f^{\rm eff}_{\rm esc,Ly}$ and $f_{\rm esc}$ likely originates mostly at the ISM-level: subsequent resonant scattering in the IGM occurs off residual HI, whose number density is affected by the value of $f_{\rm esc}$ \citep[e.g.][and Fig~4 of Dijkstra et al. 2007a for how this affects the IGM opacity]{HM12}. However, resonant scattering in the IGM depends sensitively on the assumed Ly$\alpha$ spectral line profile. Scattering off HI in galactic outflows typically redshifts the Ly$\alpha$ photons out of resonance as they escape from the galaxy, which strongly reduces the importance of resonant scattering in the IGM \citep[see Fig~1 and the discussion in \S~3.1 of][]{Dijk}. Ignoring the dependence of the IGM transmission on $f_{\rm esc}$ is therefore reasonable.\\

In short, ignoring a plausible correlation between $f^{\rm eff}_{\rm esc,Ly}$ and $f_{\rm esc}$ is a shortcoming of the model that will need to be addressed in future work. Interestingly, it may help us explain why a monotonously evolving $f_{\rm esc}$ can give rise to a non-monotonic redshift dependence of the Ly$\alpha$ fraction, with a turn-over occuring near the redshifts of interest. For example, If $f_{\rm esc}(z=3) \sim 1\%$ (consistent with upper limits by Vanzella et a. 2010) and $f_{\rm esc}(z=6) \sim 15\%$ (as in Kuhlen \& Faucher-Gigu{\`e}re 2012), then the inferred $f_{\rm esc,Ly}^{\rm eff}$ from Dijkstra \& Jeeson-Daniel (2013) and Hayes et al. (2011), implies that $y \sim 0.3-0.5$, which corresponds to $f_{\rm esc,pk} \sim 0.2-0.3$. We would have a peak in Ly$\alpha$ luminosity near the $f_{\rm esc}$ and redshift of interest.

\section{Conclusion}
\label{sec:conc}

We have investigated whether the observed reduction in Ly$\alpha$ flux from galaxies at $z>6$ can be explained by an evolving escape fraction of ionizing photons ($f_{\rm esc}$). Our study was motivated by ({\it i}) a growing consensus in the literature that $f_{\rm esc}$ must have been higher at high redshift, and ({\it ii}) the fact that it appears difficult to explain the observed reduction in Ly$\alpha$ flux with reionization alone. We found that we can reproduce the median observed drop in the Ly$\alpha$ fraction, as well as the observed evolution of the LAE luminosity functions,  with a small increase in $f_{\rm esc}$ of $\Delta f_{\rm esc} \lsim 0.1$, as long as the escape fraction is large ($f_{\rm esc} \sim0.65$) at $z\sim6$. Models with redshift evolution of $f_{\rm esc}$ that are more consistent with indirect constraints derived from observations, combined with a small evolution in global neutral fraction of $\Delta x_{\rm HI}\sim0.2$ between $z\sim7-6$ also suppress the $z\sim7$ Ly$\alpha$ flux at at the observed level. 
Our study demonstrates that an evolving escape fraction of ionising photons from galaxies modifies the observed equivalent widths of Ly$\alpha$ galaxies at a level comparable to that expected from reionization, and provides a plausible part of the explanation for evolution in the Ly$\alpha$ emission of high redshift galaxies. 

Finally, we expect the Ly$\alpha$ spectral line shape to evolve with redshift in models that invoke the IGM to explain the reduced Ly$\alpha$ flux from $z>6$ galaxies (see Fig~5 of Dijkstra et al. 2007), while this is not obviously the case for models that invoke evolution of $f_{\rm esc}$. Interestingly, \citet{Hu10} have shown that the observed Ly$\alpha$ line shape of a stack of $z\sim 6.5$ galaxies is practically identical to that at $z=5.7$. We will investigate the implications of this result in future work.

{\bf Acknowledgements} This research was conducted by the Australian Research Council Centre of Excellence for All-Sky Astrophysics (CAASTRO), through project number CE 110001020. MD acknowledges financial support from DAAD. We thank an anonymous referee for constructive comments that improved the presentation of this work.

\label{lastpage}

%\appendix
%\section{Code Testing}
%\label{app}

\end{document}